# Low Rate Wireless Personal Area Networks (LR-WPAN 802.15.4 standard)

T. Mirzoev

*The Department of Electronics and Computer Technology*
*College of Technology, Indiana State University, Terre Haute IN, 47807*

**Abstract**

This manuscript will provide a brief overview on Low Rate Wireless Personal Area Networks (LR-WPAN) for 802.15.4 protocol standard which was approved by IEEE Computer Society in May 2003. 802.15.4 standard presents some advantages for structuring sensor networks and other types of applications that require low rate communications. Some security considerations will also be presented.

## WIRELESS ADVANTAGE

Wired Local Area Networks (LAN) and other types of wired networks are struck by a massive competition from wireless technologies and networks. Major advantages of wireless networks or Wireless Local Area Networks (WLAN) are well known: wire-free networks, ease of installation, mobility and availability. In comparison to wired LANs, there are some disadvantages too: bandwidth limitations, cost of hardware and security limitations. However, despite those negative sides, wireless networks are progressively advancing in everyday life as well as providing improved standards and technologies. Speaking of standards, in May 2003, IEEE Computer Society approved 802.15.4 network protocol standard for low rate wireless devices which is another promising wireless standard.

Based on a survey, the Online Industrial Ethernet Book [2003] suggests, that a strong growth is predicted in development of industrial wireless networks. Traditional technologies such as Bluetooth and 802.11 standard will break out from their traditional applications and trend towards industrial networks. IEEE Computer Society approved 802.15.4 network protocol for low powered devices, sensor nodes and personal area networks which brings new ways for implementing wireless automation and industrial networks. This manuscript will provide an overview on low rate wireless personal area networks (LR-WPAN) and some examples of applications for 802.15.4 standard.

## ANOTHER WIRELESS STANDARD??

Variety of wireless standards such as 802.11 a/b/g followed by 802.15 (WPAN), 802.16 (Broadband Wireless Access Standards) and

P1451.5 (Wireless Sensor Standards)[1] suggests that a great need for wireless technologies exists. A simple overview of IEEE wireless standards is presented by Figure 1.

One of the standards was created due to the need for low rate wireless networks that would allow devices to have a fairly short range but have low power requirements. IEEE 802.15 Task Group 4 chairman Pat Kinney said[2] that potential uses have several things in common that they involve links that need so little power that a set of AA batteries might last three to five years or even longer.

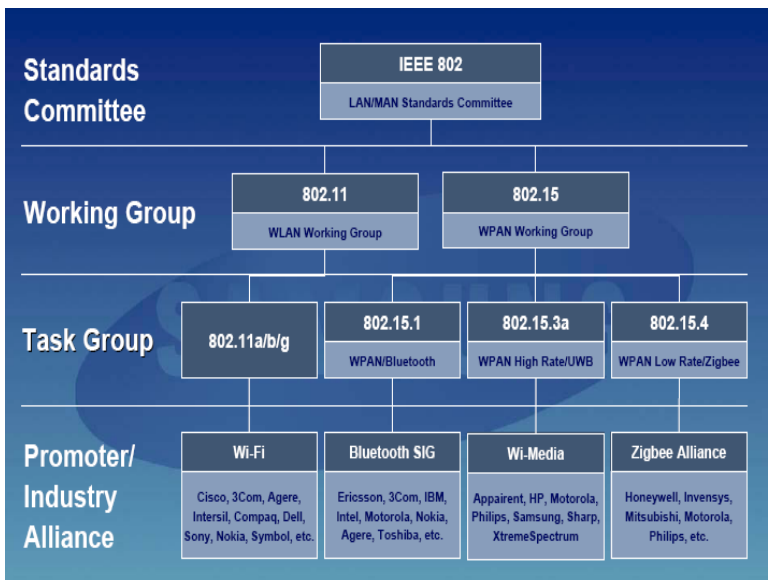

Figure 1. IEEE wireless standards

Source: Kim, T. (2004). *High Throughput Wireless Home Network Solutions*. Connectivity Lab, Digital Media R&D Center. Samsung Electronics. Retrieved on March 15, 2005 from http://snrc.stanford.edu/fileadmin/lib/Industry_Seminar/Spring_2003/tkim.pdf

"We believe a host of new applications will be based on the standard. These might include motion sensors that control lights or alarms, wall switches that can be moved at will, meter reader devices that work from outside a house, game controllers for interactive toys, tire pressure monitors in cars, passive infrared sensors for building automation systems, and asset and inventory tracking devices for use in retail stock rooms and warehouses."

802.15.4 standard targets low data rate, low power consumption and low cost wireless networking, and offers device level wireless connectivity [Zheng, Lee 2004]. Wireless personal area networks (WPAN) are designed to provide radio communications over short distances with data rates of 250 kbps, 40 kbps, and 20 kbps with two addressing modes; 16-bit short and 64-bit IEEE addressing [TG4, 2005]. In comparison to wireless local networks (WLAN), WPAN network requires little or no infrastructure [IEEE 2003]. 802.15.4 characteristics are presented by the following [IEEE 2003]:

- Over-the-air data rates of 250 kb/s, 40 kb/s, and 20 kb/s
- Star or peer-to-peer operation
- Allocated 16 bit short or 64 bit extended addresses
- Allocation of guaranteed time slots (GTSs)
- Carrier sense multiple access with collision avoidance (CSMA-CA) channel access
- Fully acknowledged protocol for transfer reliability
- Low power consumption
- Energy detection (ED)
- Link quality indication (LQI)
- 16 channels in the 2450 MHz band, 10 channels in the 915 MHz band, and 1 channel in the 868 MHz band

802.15.4 architecture is presented by Figure 2.

## MAKING CONNECTIONS

In order to create a network, an understanding of simple connection nodes and a basic structure is needed. For LR-WPAN networks those nodes and structure is presented by the possible topologies on Figure 3. 802.15.4 protocol allows for the following network topologies: 1) Star Topology, 2) Peer-To-Peer Topology, 3) Combined Topology.

---

[1] IEEE Computer Society website, *IEEE Wireless Standards Zone*. Retrieved from on April 1, 2005 form http://standards.ieee.org/wireless/index.html

[2] The Online Industrial Ethernet Book, *Wireless News*. Retrieved on February 14, 2005 from http://ethernet.industrial-networking.com/wireless/news.asp#122

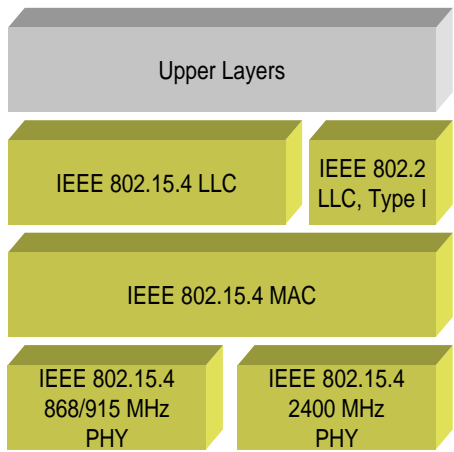

Figure 2. 802.15.4 Architecture

Figure 3. shows some devices that are specific to LR-WPAN networks. For example, there are two device classes that can participate in an 802.15.4 network [IEEE 15-04-0218-00-004a]:
1. Full function device (FFD)
   - Any topology
   - PAN coordinator capable
   - Talks to any other device
   - Implements complete protocol set
2. Reduced function device (RFD)
   - Limited to star topology or end-device in a peer-to-peer network.
   - Cannot become a PAN coordinator
   - Very simple implementation
   - Reduced protocol set

The above device classifications introduce some new definitions, such as [IEEE 15-04-0218-00-004a]:
- *Network Device:* An RFD or FFD implementation containing an IEEE 802.15.4 medium access control and physical interface to the wireless medium.
- *Coordinator:* An FFD with network device functionality that provides coordination and other services to the network.
- *PAN Coordinator:* A coordinator that is the principal controller of the PAN. A network has exactly one PAN coordinator.

Depending on a type of topology that is used for LR-WPAN network, devices class should be chosen accordingly. In other words, by looking at the Figure 4. it is noticeable, that a different number of Full Function and Reduced Function Devices is used depending on the application/network requirements.

Besides the advantages of LR-WPAN networks, there are some concerns and precautions exist for those that study or develop LR-WPAN. Especially, when sensor networks are being constructed based on 802.15.4 standard. Sastry [2004] and Wagner [2004] suggest certain flaws exist in Key Management, Integrity Protection of the Media Access Control Security Suite of 802.15.4 protocol that create danger zones for wireless industrial networks and application developers. Consequently, a detailed research of security flaws of 802.15.4 wireless networks should allow scientists to avoid future possible errors in industrial wireless networks or even in critical infrastructure cyber security policy creations.

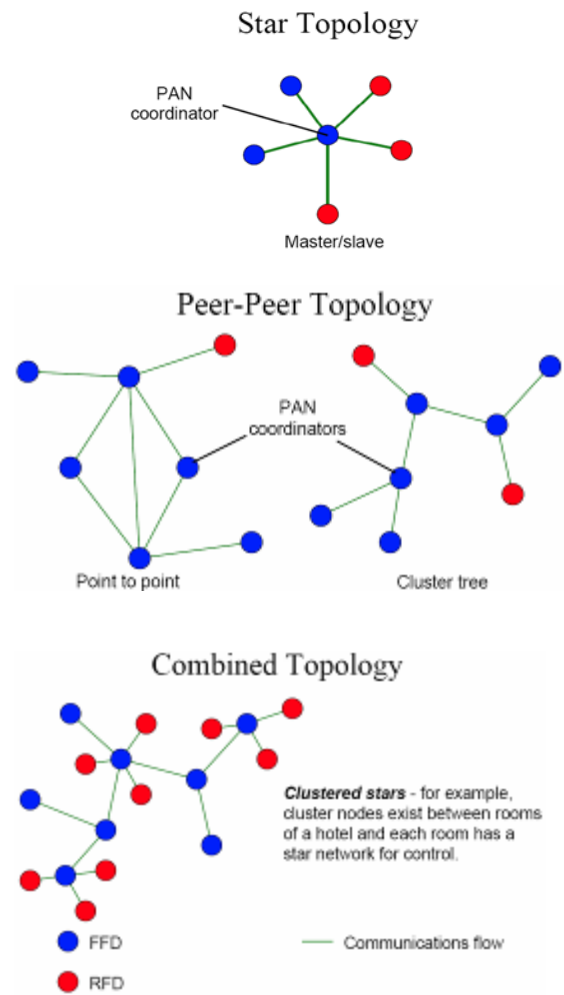

Figure 3. 802.15.4 Topologies

## CONCLUSION

In conclusion, it is important to summarize this overview of LR-WPAN networks. 802.15.4 standard creates new advantages for the world of wireless communication. Like any other protocol, it has its advantages and disadvantages. IEEE 802.15 Task Group 4 chairman Pat Kinney said that main applications of LR-WPAN include the following[3]:

- motion sensors that control lights or alarms
- wall switches that can be moved at will,
- meter reader devices that work from outside a house
- game controllers for interactive toys
- tire pressure monitors in cars
- passive infrared sensors for building automation systems
- asset and inventory tracking devices for use in retail stock rooms and warehouses

802.15.4 is a promising standard, but like other newly introduced standards it needs improvements in some of its parts. Along with the industrial development of the LR-WPAN applications, the necessary improvements might be suggested and implemented.

## BIBLIOGRAPHY

GossamerNet (2003). Wireless Development Environment: Technology. *Copyright Red Wing Technologies, Inc., 2003*. Retrieved on March 30, 2005 from http://www.gossamernet.com/lrwpan/technology.asp

IEEE 15-04-0218-00-004a document, *Project: IEEE P802.15 Working Group for Wireless Personal Area Networks (WPANs)*. Retrieved on March 10 2005 from http://www.ieee802.org/15/pub/TG4Expert.html

IEEE Computer Society (2003). Part 15.4: *Wireless Medium Access Control (MAC) and Physical Layer (PHY) Specifications for Low-Rate Wireless Personal Area Networks (LR-WPANs)*. Tech. Rep.

Kim, T. (2004). *High Throughput Wireless Home Network Solutions*. Connectivity Lab, Digital Media R&D Center. Samsung Electronics. Retrieved on March 15, 2005 from http://snrc.stanford.edu/fileadmin/lib/Industry_Seminar/Spring_2003/tkim.pdf

Oman P., Schweitzer E. III, Roberts J., (2001). *Safeguarding IEDS, Substations, and SCADA systems against electronic intrusions*. Schweitzer Engineering Laboratories, Inc. 2001

Pollet, J (2002). *SCADA Security Strategy*, PlantData Technologies.

Sastry, N., Wagner D. (2004). *Security Considerations for IEEE 802.15.4 Networks*. WiSE'04, Philadelphia, Pennsylvania.

Schellekens, P. (2001). *A Vision for Intelligent Instrumentation in Process Control*. Control Engineering Online, October 2001.

The Online Industrial Ethernet Book, *Wireless News*. Retrieved on February 14, 2005 from http://ethernet.industrial-networking.com/wireless/news.asp#122

WPAN™ Task Group 4 (TG4) website (2005). IEEE 802.15 Working Group for WPAN IEEE 802.15. Retrieved on March 13, 2005 from http://www.ieee802.org/15/pub/TG4.html

Zeng J., Lee M. (2004). *A Comprehensive Performance Study of IEEE 802.15.4*. The City University of New York, NY, Department of Electrical Engineering.

---

[3] The Online Industrial Ethernet Book, *Wireless News*. Retrieved on February 14, 2005 from http://ethernet.industrial-networking.com/wireless/news.asp#122